\documentclass[12pt,preprint]{aastex}

\def\la{\mathrel{\hbox{\rlap{\hbox{\lower4pt\hbox{$\sim$}}}\hbox{$<$}}}}
\def\sun{\hbox{$\odot$}}

\begin{document}

\title{Transition redshift: new kinematic constraints from supernovae}
\author{J. V. Cunha}
\email{cunhajv@astro.iag.usp.br}
\author{J. A. S. Lima}
\email{limajas@astro.iag.usp.br}
\affil{Departamento de Astronomia,
Universidade de S\~ao Paulo, USP,
\\ 05508-900 S\~ao Paulo, SP, Brazil}
\begin{abstract}
The transition redshift (deceleration/acceleration) is discussed  by
expanding the deceleration parameter to first order around its
present value. A detailed study is carried out by considering two
different parameterizations: $q=q_0 + q_1z$ and $q=q_0 + q_1
z(1+z)^{-1}$, and the associated free parameters ($q_o, q_1$) are
constrained by 3 different supernova samples. The previous analysis
by Riess {\it{et al.}} [ApJ 607, 665, 2004] using the first
expansion is slightly improved and confirmed in light of their
recent data ({\emph{Gold}}07 sample). However, by fitting the model
with the Supernova Legacy Survey (SNLS) type Ia sample we find that
the best fit to the redshift transition is $z_t = 0.61$ instead of
$z_t = 0.46$ as derived by the High-z Supernovae Search (HZSNS)
team. This result based in the SNLS sample is also in good agreement
with  the Davis {\it{et al.}} sample, $z_t=0.60^{+0.28}_{-0.11}$
($1\sigma$). Such results are in line with some independent analyzes
and accommodates more easily the concordance flat model
($\Lambda$CDM). For both parameterizations, the three SNe type Ia
samples considered favor recent acceleration and past deceleration
with a high degree of statistical confidence level. All the
kinematic results presented here depend neither on the validity of
general relativity nor the matter-energy contents of the Universe.
\end{abstract}

\section{INTRODUCTION}
\label{sec:intro}
It is now widely believed that the universe at redshifts smaller
than unity underwent a ``dynamic phase transition" from decelerating
to accelerating expansion which has been corroborated by several
independent analyzes. In the context of the general relativity
theory such a phenomenon can be interpreted as a dynamic influence
of some sort of dark energy whose main effect is to change the sign
of the universal decelerating parameter $q(z)$.

The most direct observation supporting the present accelerating
stage of the Universe comes from the luminosity distance versus
redshift relation measurements using supernovae (SNe) type Ia (Riess
{\it et al.} 1998; Perlmutter {\it et al.} 1999; Riess {\it et al.}
2004, 2007; Astier {\it{et al.}} 2006; Wood-Vasey {\it et al.} 2007;
Davis {\it et al.} 2007; Kowalski M., {\it et al.} 2008) initially
interpreted in light of $\Lambda$CDM scenarios using either
background or inhomogeneous luminosity distances (Santos {\it {et
al.}} 2008; Santos \& Lima 2008). However, independent theoretical
observational analyzes points to more general models whose basic
ingredient is a negative-pressure dark energy component (for review
see, Padmanabhan 2003; Peebles \& Ratra 2003; Lima 2004; Copeland
{\it{et al.}} 2006).

The convergence of many high-quality experimental results are now
strongly suggesting that the cosmic picture at present is the
following: the spatial geometry (curvature) is flat, or more
approximately flat, and the dynamics is governed by a component
called dark energy, 3/4 of composition, and 1/4 for matter component
(baryons plus dark). Among a number of possibilities to describe
this dark energy component, the simplest and most theoretically
appealing way is by means of a positive cosmological constant
$\Lambda$. Others possible candidates are:  a vacuum decaying energy
density, or a time varying $\Lambda$-term (Ozer \& Taha 1987; Freese
{\it{et al.}} 1987; Carvalho {\it{et al.}} 1992; Lima \& Maia 1993;
1994; Maia \& Lima 1999; 2002; Lima 1996; Overduin \& Cooperstock
1998; Cunha {\it{et al.}} 2002a,b; Cunha \& Santos 2004; Alcaniz \&
Lima 2005; Carneiro \& Lima 2005; Fabris {\it {et al.}} 2007), a
time varying relic scalar field slowly rolling down its potential
(Ratra \& Peebles 1988; Frieman {\it{et al.}} 1995; Caldwell {\it{et
al.}} 1998; Saini {\it{et al.}} 2000; Caldwell 2000; Carvalho {\it
{et al.}} 2006), the so-called ``X-matter", an extra component
simply characterized by an equation of state $p_{\rm x}=
\omega\rho_{\rm x}$ (Turner \& White 1997; Chiba {\it{et al.}} 1997;
Alcaniz \& Lima 1999, 2001; Kujat {\it{et al.}} 2002; Alcaniz
{\it{et al.}} 2003), the Chaplygin gas whose equation of state is
given by $p= -A/\rho$ where $A$ is a positive constant (Kamenshchik
{\it{et al.}} 2001; Bento {\it{et al.}} 2002; Dev {\it{et al.}}
2003; Bento {\it{et al.}} 2003; Cunha {\it{et al.}} 2004; Lima
{\it{et al.}} 2006a,b), among others (Lima \& Alcaniz 1999; Chimento
{\it{et al.}} 2001; Freese \& Lewis 2002; Pavon \& Zimdahl 2005).
For SFC and XCDM scenarios, the $\omega$ parameter may be a function
of the redshift (see, for example, Efstathiou 1999; Cunha {\it et
al.} 2007), or still, as it has been recently discussed, it may
violate the dominant energy condition and assume values $<-1$ when
the extra component is named phantom cosmology (Caldwell 2002; Lima
{\it{et al.}} 2003; Perivolaropoulos 2005; Gonzalez-Diaz \& Siguenza
2004; Lima \& Alcaniz 2004; Santos \& Lima 2008). It should be
stressed, however, that all these models are based on the validity
of general relativity or some of its scalar-tensorial
generalizations.

On the other hand, Turner \& Riess (2002) have discussed an
alternative route - sometimes called kinematic approach - in order
to obtain information about the beginning of the present
accelerating stage of the Universe with no assumption concerning the
validity of general relativity or even of any particular metric
gravitational theory (in this connection see also Weinberg 1972).
Although considering that such a method does not shed light on the
physical or geometrical properties of the new energetic component
causing the acceleration it allows one to assess the direct
empirical evidence for the transition from decelerating to
accelerating in the past as provided by SNe type Ia measurements. In
their preliminary analysis it was found that the SNe data favor
recent acceleration ($z < 0.5$) and past deceleration ($z > 0.5$).

More recently, with basis on the same approach, the High-z Supernova
Search (HZSNS) team have obtained $z_t = 0.46 \pm 0.13$ at $1\sigma$
c.l. (Riess {\it{et al.}} 2004) which has been further improved to
$z_t = 0.43 \pm 0.07$ ($1\sigma$) c.l. (Riess {\it et al.} 2007).
Many authors have used these values of the transition redshift for
imposing constraints on the cosmic parameters as an independent and
trustworth discriminator for cosmology (Gardner 2005; Virey {\it{et
al.}} 2005; Qiang \& Zhang 2006; Gong \& Wang 2006; Gong \& Wang
2007, Xu {\it{et al.}} 2007; Ishida {\it{et al.}} 2008).

In this article, we constrain the transition redshift by expanding
the deceleration parameter around its present value. A detailed
analysis is carried out for 2 different parameterizations: $q=q_0 +
q_1z$ and $q=q_0 + q_1 z(1+z)^{-1}$. We show that the kinematic
analysis based on the SNLS ({Astier {\it{et al.}} 2006}), as well as
the one  recently compiled by Davis {\it{et al.}} 2007, yield a
transition redshift $z_t\sim 0.6$ ($1\sigma$) in better agreement
with the flat $\Lambda$CDM model {\bf ($z_t =
(2\Omega_{\Lambda}/\Omega_m)^{1/3} -1 \sim 0.66$)} than the one
provided by Riess {\it {et al.}} (2007) using the High-z Supernova
Search (HZSNS) sample.

\section{KINEMATIC APPROACH AND ANALYSIS}

\begin{figure}[ht]
\centering
\includegraphics[width=120mm, angle=0]{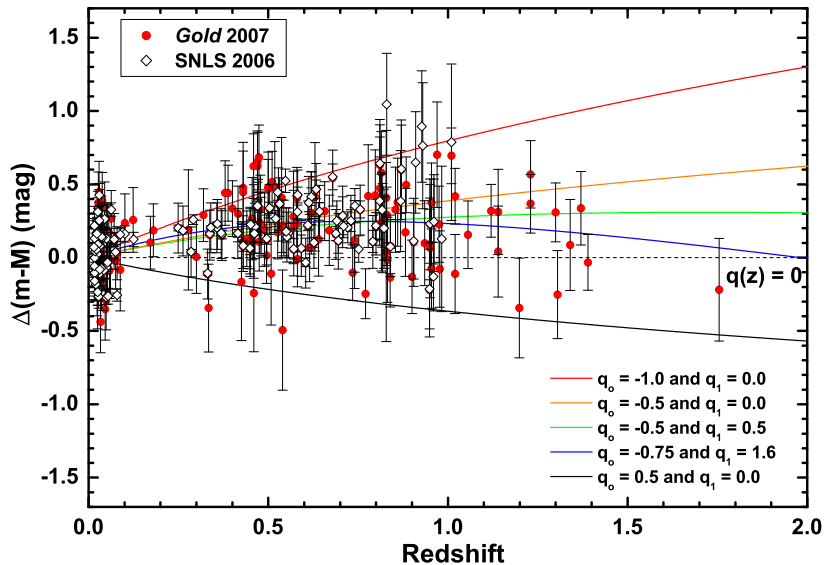}
\caption{Residual magnitude  versus redshift is shown for SNe type
Ia. The SNe of Riess {\it {et al.}} 2007 are represented by red
circles, and, the remaining ones (Astier {\it {et al.}} 2006, SNLS)
by open diamonds. Data and kinematic models of the expansion history
are shown relative to an eternally coasting model, $q(z)= 0$. Models
representing specific kinematic scenarios are illustrated.}
\label{Fig1}
\end{figure}

To begin with, let us assume that the Universe is spatially flat as
motivated by inflation and the WMAP results (Spergel {\it {et al.}}
2007, Komatsu et al. 2008). Following Turner \& Riess (2002) and
Riess {\it{et al.}} (2004), the luminosity distance is defined by
the following integral expression (with $c=1$)

\begin{eqnarray}\label{eq:dLq}
D_L(z) = (1+z)\int_0^z {du\over H(u)} \hspace{3.cm} & & \nonumber \\
= \frac{(1+z)}{H_0} \int_0^z \exp{\left[-\int_0^u [1+q(u)]d\ln
(1+u)\right]} du, &
\end{eqnarray}
where the $H(z)=\dot a/a$ is the Hubble parameter, and, $q(z)$, the
deceleration parameter, is defined by
\begin{eqnarray}\label{qz}
q(z)\equiv -\frac{a\ddot a}{\dot a^2} = \frac{d H^{-1}(z)}{ dt} -1.
\end{eqnarray}

Although generalizable for non-zero curvature,  Eq. (1) is not a
crude approximation as one may think at first sight.  In the
framework of a flat FRW type universe, it is an exact expression for
the luminosity distance which depends on the expressions of the
present Hubble constant, $H_0$, and the epoch-dependent deceleration
parameter, $q(z)$.  The simplest way to work with the coupled
definitions (1) and (2) as a kinematic  model for the SN type Ia
data is by adopting a parametric representations for $q(z)$. As one
may check, in the case of a linear two-parameter expansion for
$q(z)=q_o +z{q_1}$ (Riess {\it{et al.}} 2004), the integral (1) can
be represented in terms of a special function as (see Appendix A)
\begin{eqnarray}\label{eq:dLKin}
D_L(z) &=& \frac{(1+z)}{H_0}e^{q_1}q_1{^{q_o-q_1}}
[\gamma{({q_1-q_o},(z+1){q_1})} \nonumber \\
&& \,\, - \gamma{({q_1-q_o},{q_1})}],
\end{eqnarray}
where ${q_o}=q(z=0)$ is the present value  of the deceleration
parameter, ${q_1}$ is the derivative in the redshift evaluated at
$z=0$, and $\gamma$ is the incomplete gamma function (see,
Abramowitz \& Stegun 1972) with the condition ${q_1-q_o}>0$ must be
satisfied (more details in the Appendix). By using the above
expressions we  may get information about $q_o$, $q_1$ and,
therefore, about the global behavior of $q(z)$. Note also that a
positive transition redshift, $z_t$, may be obtained only for
positive signs of $q_1$ (the variation rate of $q_o$) since $q_o$ is
negative and the dynamic transition (from decelerating to
accelerating) happens at $q(z_t)=0$, or equivalently,
$z_t=-q_o/q_1$.

\begin{figure}[ht]
\centering
\includegraphics[width=80mm, angle=0]{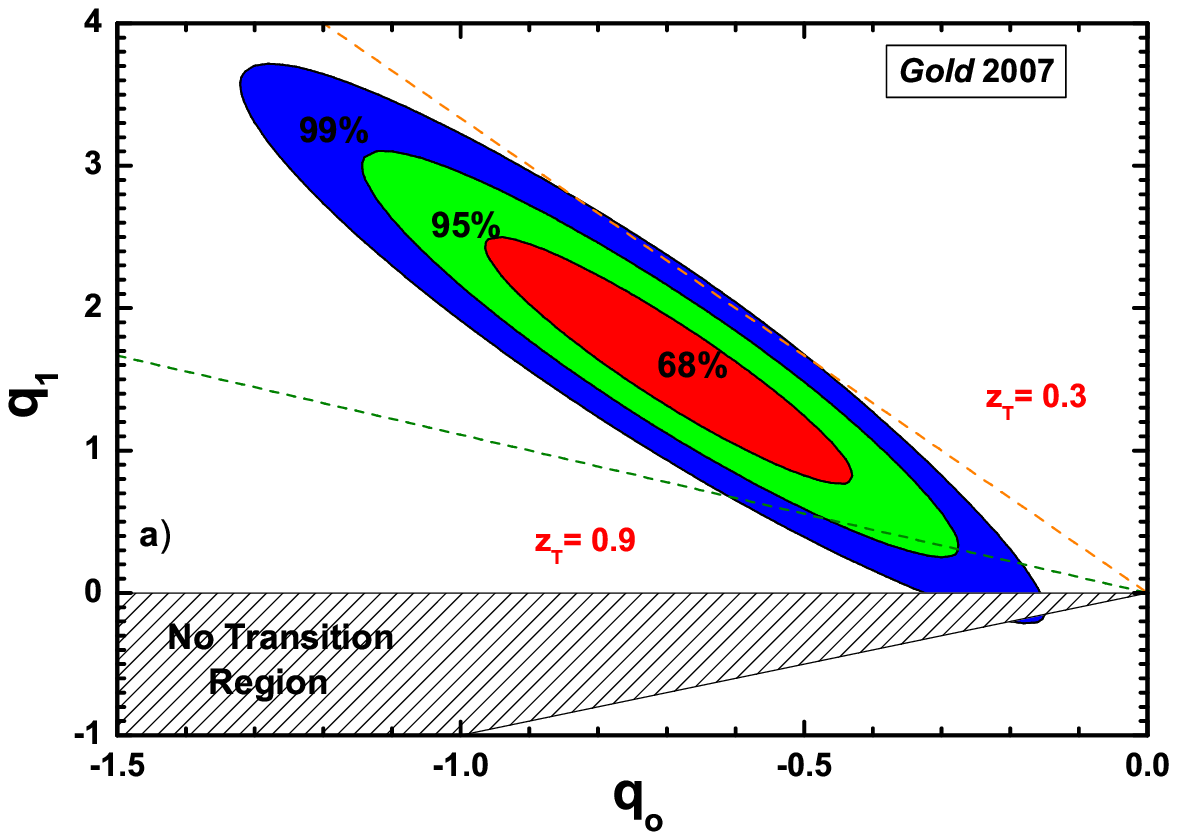}
\includegraphics[width=80mm, angle=0]{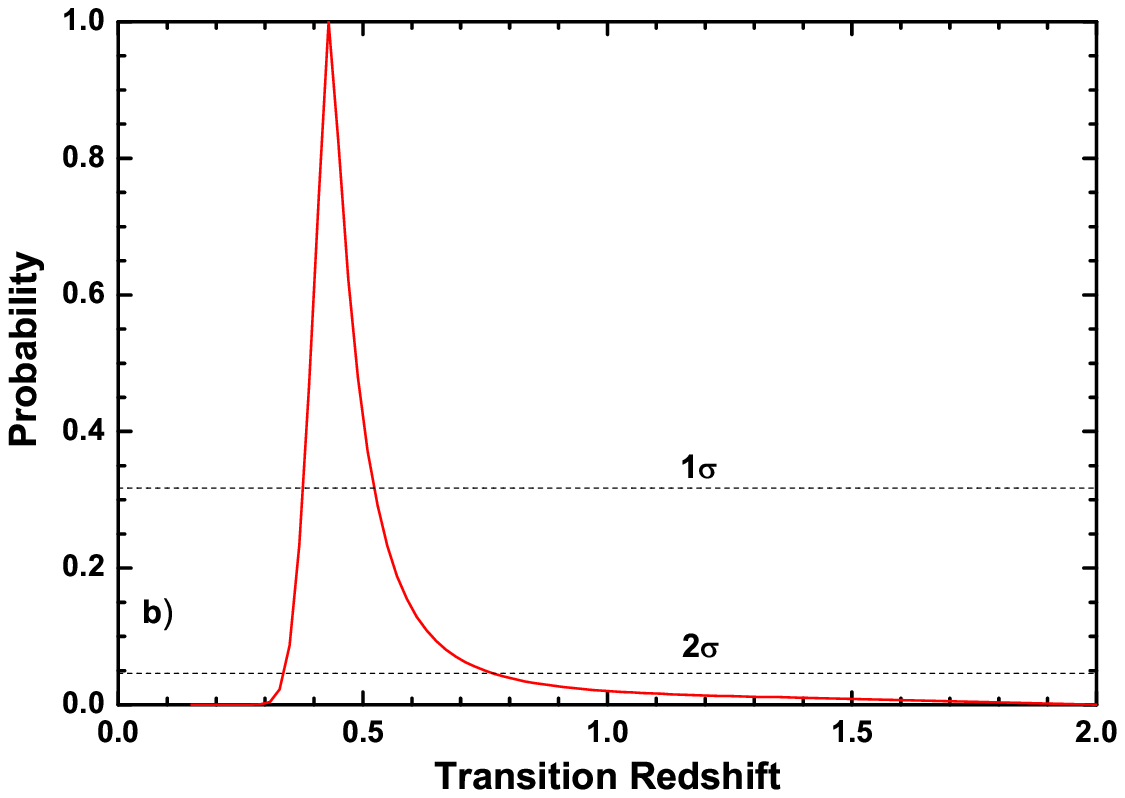}
\caption{{\bf a)} Contours in the $q_o - q_1$ plane from 182 SNe of
Riess {\it {et al.}} (2007).  The two-parameter model of the
expansion history is $q(z)=q_o+q_1z$, and the contours correspond to
68.3\%, 95.4\% and 99.7\% c.l. as indicated. The best fit to the
pair of free parameters is ($q_o,q_1$) = ($-0.7,1.62)$. {\bf b)}
Likelihood function for the past transition redshift. The best fit
is $z_t = 0.43^{+0.09}_{-0.05}$ ($1\sigma$) $^{+0.29}_{-0.09}$
($2\sigma$).} \label{Fig2}
\end{figure}

Another parametrization of considerable interest is $q(z)=q_o + q_1
z/(1+z)$ (Xu {\it{et al.}} 2007). It has the advantage to be well
behaved at high redshift while the linear approach diverges at the
distant past. In this case we can write the integral (1) as (see
Appendix A)
\begin{eqnarray}\label{eq:dLKin2Parame}
D_L (z) &=& \frac{(1+z)}{H_{0}}e^{q_{1}}{q_1^{-(q_o+q_1)}}
[\gamma(q_{1}+q_{o},q_1) \nonumber \\
& & \,\, - \gamma{({q_1+q_o},q_1/(1+z))}],
\end{eqnarray}
where ${q_o}=q(z=0)$ is the present value  of the deceleration
parameter, ${q_1}$ is the parameter yielding the total correction in
the distant past ($z\gg 0, q(z)= q_o + q_1$). Note also that a
positive transition redshift, $z_t$, may be obtained only for
positive signs of $q_1 > |q_o|$  since $q_o$ is negative and the
dynamic transition (from decelerating to accelerating) happens at
$q(z_t)=0$, or equivalently, $z_t=-q_o/(q_o+q_1)$.

The so-called \emph{Gold}07 sample from the HZSNS team is a
selection of 182 SNe Ia events distributed over the redshift
interval $0.001 < z < 1.8$, and constitutes the compilation of the
best observations made so far by them which were completed by  16
events from the Supernova Cosmology Project observed with the Hubble
Space Telescope (HST).

The current data from SNLS collaboration correspond to the first
year results of its planned five year survey. The total sample
includes 71 high-$z$ SNe Ia in the redshift range $0.2 < z < 1$ plus
44 low-$z$ SNe Ia. This data set is arguably (due to multi-band,
rolling search technique and careful calibration) the best high-$z$
SNe Ia compilation to date, as indicated by the very tight scatter
around the best fit in the Hubble diagram and a careful estimate of
systematic uncertainties. Another important aspect to be emphasized
on the SNLS data is that they seem to be in a better agreement with
WMAP 3-years results (Spergel {\it et al.} 2007) than the previous
\emph{Gold04} sample observed by Riess {\it et  al.} (2004). For a
more detailed discussion see e.g., Jassal {\it{et al.}} 2006.

For completeness, we also consider the Type Ia Supernovae
compilation by Davis {\it{et al.}} 2007. This extended sample is
formed by 192 SNIa consisting of 45 SNe from a nearby SNe Ia
subsample, plus 57 from SNLS and 60 intermediate redshift SNe from
ESSENCE (Wood-Vasey {\it{et al.}} 2007), and, finally, 30 high
redshift \emph{Gold07} SNe with internally consistent magnitude
offsets. The supernovae span the redshift range $0.2 < z < 1.8$
(Riess {\it{et al.}} 2007).

\begin{figure}[ht]
\centering
\includegraphics[width=80mm, angle=0]{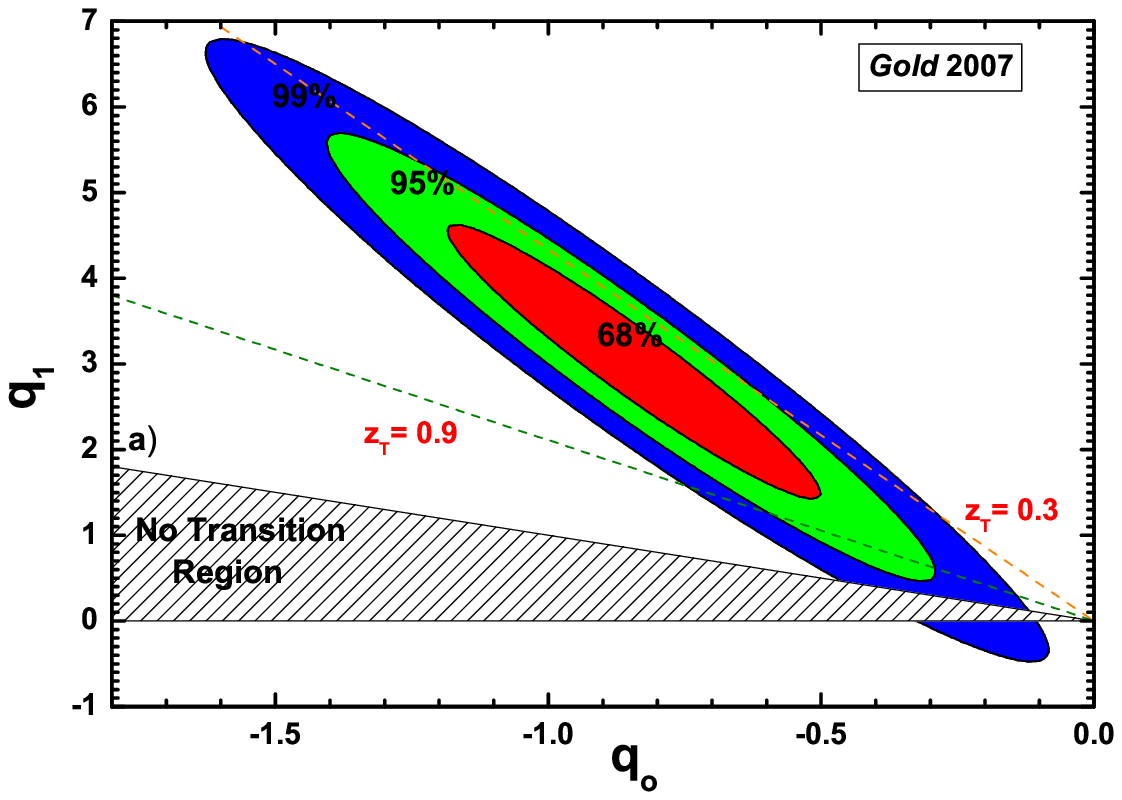}
\includegraphics[width=80mm, angle=0]{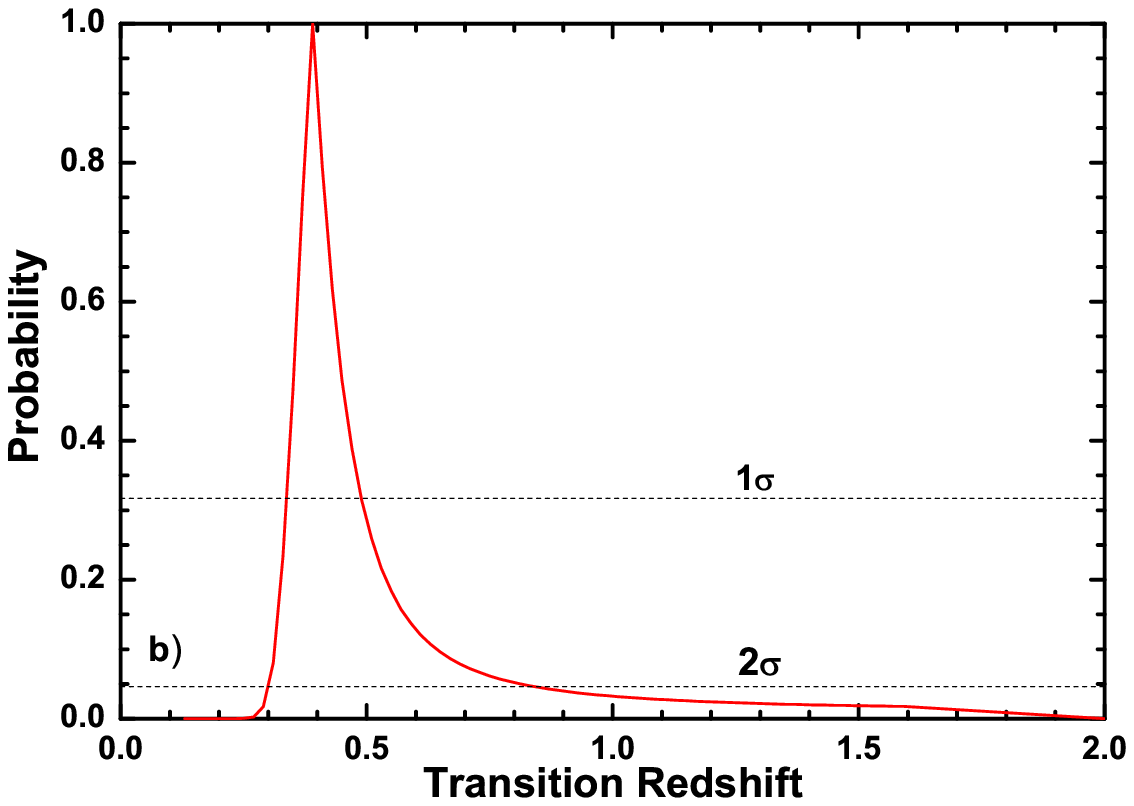}
\caption{{\bf a)} Contours in the $q_o - q_1$ plane for 182 SNe type
Ia data ({\bf a)} Contours in the $q_o - q_1$ plane for 182 SNe type
Ia data (\emph{Gold}07 sample) considering $q(z)=q_o +q_1z/(1+z)$.
Now, the best fit to the pair of free parameters is ($q_o,q_1$) =
($-0.84,3)$. {\bf b)} Likelihood function for the past transition
redshift. The best fit is $z_t = 0.39^{+0.10}_{-0.06}$ ($1\sigma$)
$^{+0.44}_{-0.09}$ ($2\sigma$).} \label{Fig3}
\end{figure}

In figure 1 we show the theoretical predictions of the kinematic
approach to the residual Hubble diagram with respect to an eternally
coasting Universe model ($q(z)\equiv0$). Such a diagram is usually
presented in terms of the bolometric distance modulus which is
defined by $m-M = 5 logD_{L}(z) + 25$ (Peebles 1993). The different
models are characterized by the selected values of $q_o$ and $q_1$,
as depicted in the diagram. The SNe type Ia data shown in the panel
comprise both the SNLS (Astier {\it {et al.}} 2006) and HZSNS (Riess
{\it {et al.}} 2007) samples as indicated. At this point, it is
natural to ask about the likelihood contours in the plane
($q_o,q_1$) and the probability of the transition redshift derived
for each sample separately, as well as to the whole set of data.

In order to obtain that, let us consider the maximum likelihood that
can be determined from a $\chi^2$ statistics for a given set of
parameters $\mathbf{p}$
\begin{equation}
\chi^2(z|\mathbf{p}) = \sum_i { (\mu_{p,i}(z_i;
\mathbf{p})-\mu_{0,i})^2 \over \sigma_{\mu_{0,i}}^2+\sigma_v^2},
\end{equation}
where $\sigma_{\mu_{0,i}}$ is the uncertainty in the individual
distance moduli, $\sigma_{v}$ is the  dispersion in SNe redshifts
due to peculiar velocities (Riess {\it et al.} 2004, 2007), and the
complete set of parameters is $\mathbf{p} \equiv (H_o, q_o, q_1)$.

In what follows we investigate the bounds arising on the empirical
$q(z)$ parameters and the probability of the redshift transition for
each SNe type Ia sample. By marginalizing the likelihood function
over the nuisance parameter, $H_0$, the contours and the
probabilities of the transition redshift for each sample are readily
computed.

\subsection{Gold Sample Analysis:  $q=q_o + q_1{z}$}

In figure 2a we show the plane $q_o-q_1$ for confidence levels of
$68.3\%$, $95.4\%$ and $99.7\%$ for the \emph{Gold}07 sample. Note
that it strongly favors a Universe with recent acceleration
($q_o<0$) and previous deceleration ($dq/dz>0$). With one free
parameter the confidence region is $-1.06 \leq q_o \leq -0.42$ and
$0.6 \leq q_1 \leq 2.8$ with ($95.4\%$) confidence level. It should
be remarked the presence of a forbidden region forming a trapezium.
The horizontal line in the top is defined by $q_1=0$ which leads to
an infinite (positive or negative) transition redshift while the
segment at $45^{o}$ is the infinite future ($z_t = -1$). Actually,
since $z=a_o/a -1$, in the infinite future ($a \rightarrow \infty$)
one finds that $z \rightarrow -1$. In addition, the values of $z_t$
associated with the horizontal segment in the bottom are always
smaller than -1 (therefore, after the infinite future!) since $-1.5
\leq z_t \leq -1$. Finally,  one may conclude that the vertical
segment is associated with $z_t \leq -1.5$, thereby demonstrating
that the hachured trapezium is actually a physically forbidden
region.

In Figure 2b, one may see the probability of the associated
transition redshift $z_t$ defined by $q(z_t)=0$. It has been derived
by summing the probability density in the $q_0$ versus the ${dq/dz}$
plane along lines of constant transition redshift, $z_t =
-q_0/(dq/dz)$. The resulting analysis yields $z_t =
0.43^{+0.09}_{-0.05}$ ($1\sigma$) $^{+0.29}_{-0.09}$ ($2\sigma$) for
one free parameter which is in reasonable agreement with the value
$z_t = 0.46 \pm 0.13$ quoted by Riess {\it et al.} (2004).  In our
analysis, the asymmetry in the probability of $z_t$ is produced by a
partially parabolic curve obtained when $\chi^2$ is minimized. It is
clear that the central value ($z_t=0.43$) does not agree with
several dynamical flat models in $68.3$\% and $94.5$\% confidence
level. In particular, $z_t= 0.75$ in the flat $\Lambda$CDM
concordance model ($\Omega_m= 0.27$).

\subsection{Gold Sample:  $q=q_o + q_1{z}/{1 + z}$}

In figure 3a we show the plane $q_o-q_1$ for confidence levels of
$68.3\%$, $95.4\%$ and $99.7\%$ using the \emph{Gold}07 sample for
the above phenomenological law. Again, we see that a Universe with
recent acceleration ($q_o<0$) and previous deceleration ($dq/dz>0$)
is strongly favored. With one free parameter the allowed region is
$-1.25 \leq q_o \leq -0.15$ and $-1.5 \leq q_1 \leq 4.9$ with
($95.4\%$) confidence level.  Note also the presence of a physically
forbidden region forming a trapezium (cf Fig. 2a). As one may check,
the horizontal line at the bottom is defined by $q_1=0$ which leads
to a transition redshift in the infinite future ($z_t = -1$) while
the  $45^{o}$ segment also defines an extreme line of transition
redshifts. Note also that all transition redshifts associated with
the vertical segment of the triangle (on the axis defined by $q_1$)
take values on the future, more precisely, on the interval
[-1,-1/2].

In Figure 3b we display the probability of the transition redshift
$z_t$ which is defined by $q(z_t)=0$. The resulting analysis yields
$z_t = 0.39^{+0.10}_{-0.06}$ ($1\sigma$) $^{+0.44}_{-0.09}$
($2\sigma$) for one free parameter. This central value agrees with
the flat concordance model only at $95.4$\% ($2\sigma$). Note that
the straight lines denoting $z_t= 0.3$ (transition redshift for a
flat $\Lambda$CDM with $\Omega_m\simeq \Omega_{\Lambda}\simeq 0.5$)
and $z_t= 0.9$ (for a $\Lambda$CDM with $\Omega_m\simeq 0.2$ and
$\Omega_{\Lambda}\simeq 0.8$) are outside of the $2\sigma$ regions
in both parametrizations (see Figs. 2a and 3a).

\begin{figure}[ht]
\centering
\includegraphics[width=80mm, angle=0]{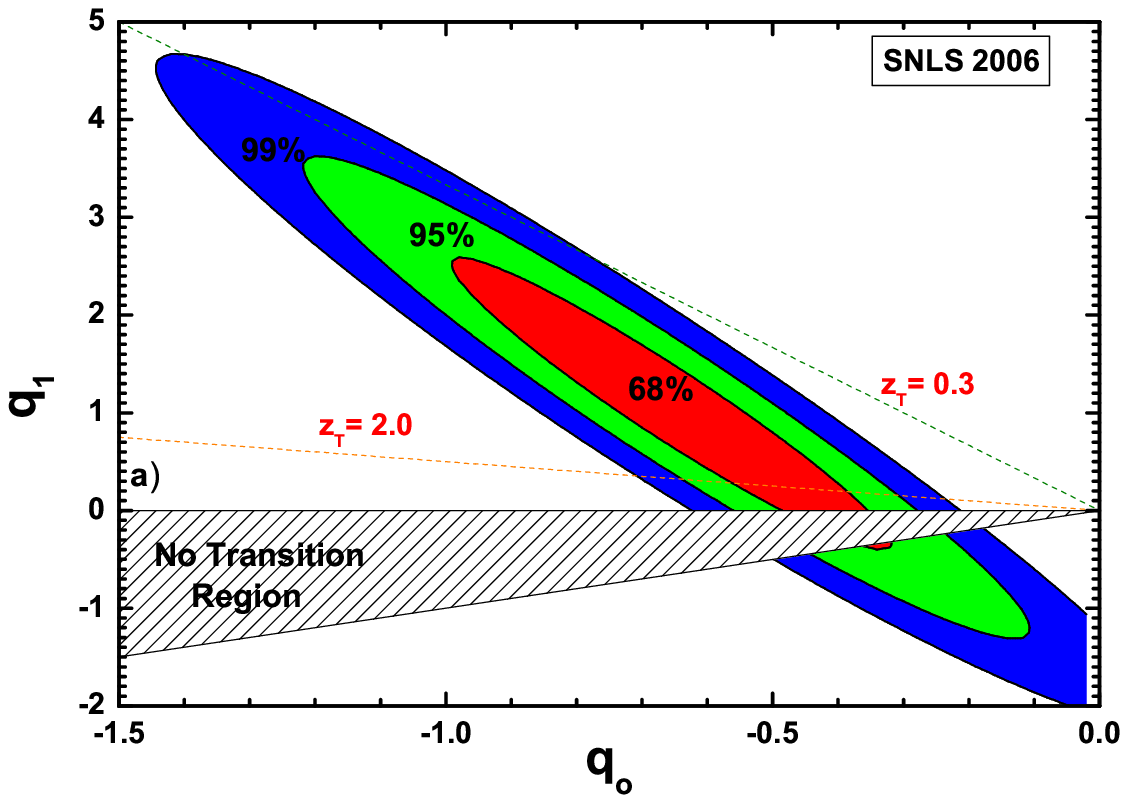}
\includegraphics[width=80mm, angle=0]{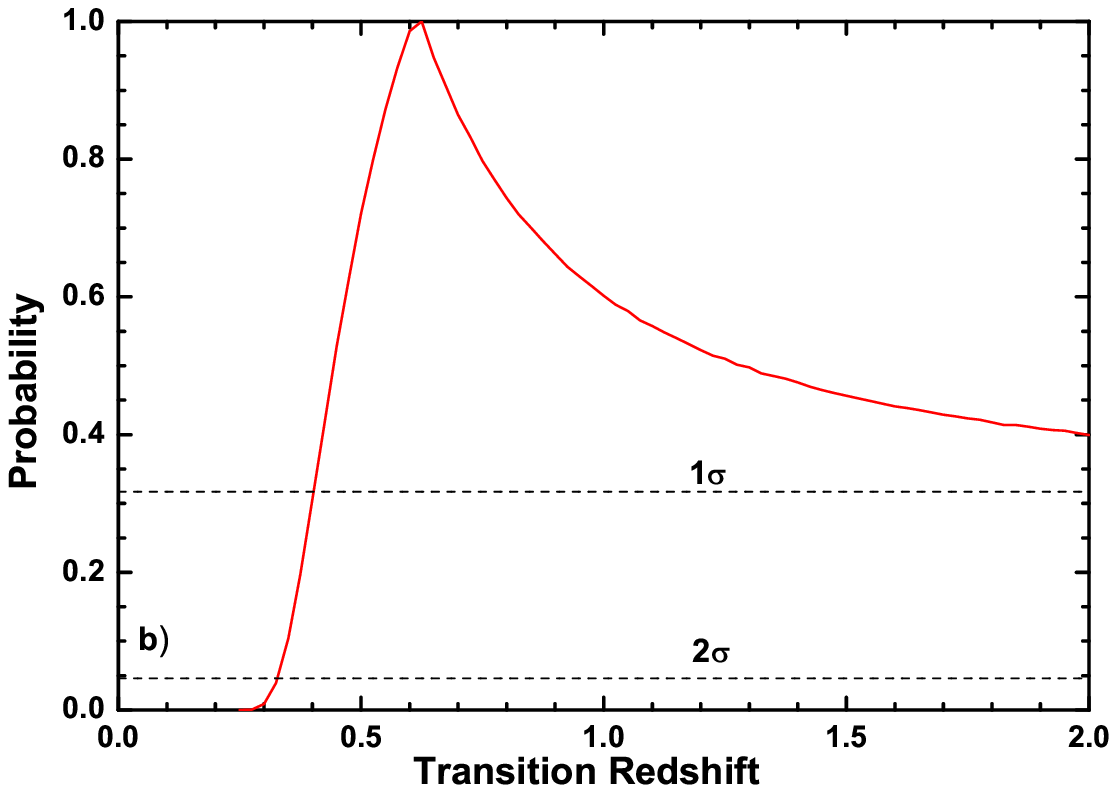}
\caption{{\bf a)} Constraints on the deceleration parameter from the
SNLS sample with law $q(z)=q_o+q_1z$. Data favor the recent
acceleration ($q_0<0$) and past deceleration ($q_1>0$). Confidence
intervals are showed in the diagram. {\bf b)} Likelihood function
for the past transition redshift. Due to the absence of data at high
redshifts, we see that peak of the likelihood function for the SNLS
sample is still pronounced but the upper limit of the transition
redshift is not very well defined (confront with Fig. 2b).}
\label{Fig4}
\end{figure}

\begin{figure}[ht]
\centering
\includegraphics[width=80mm, angle=0]{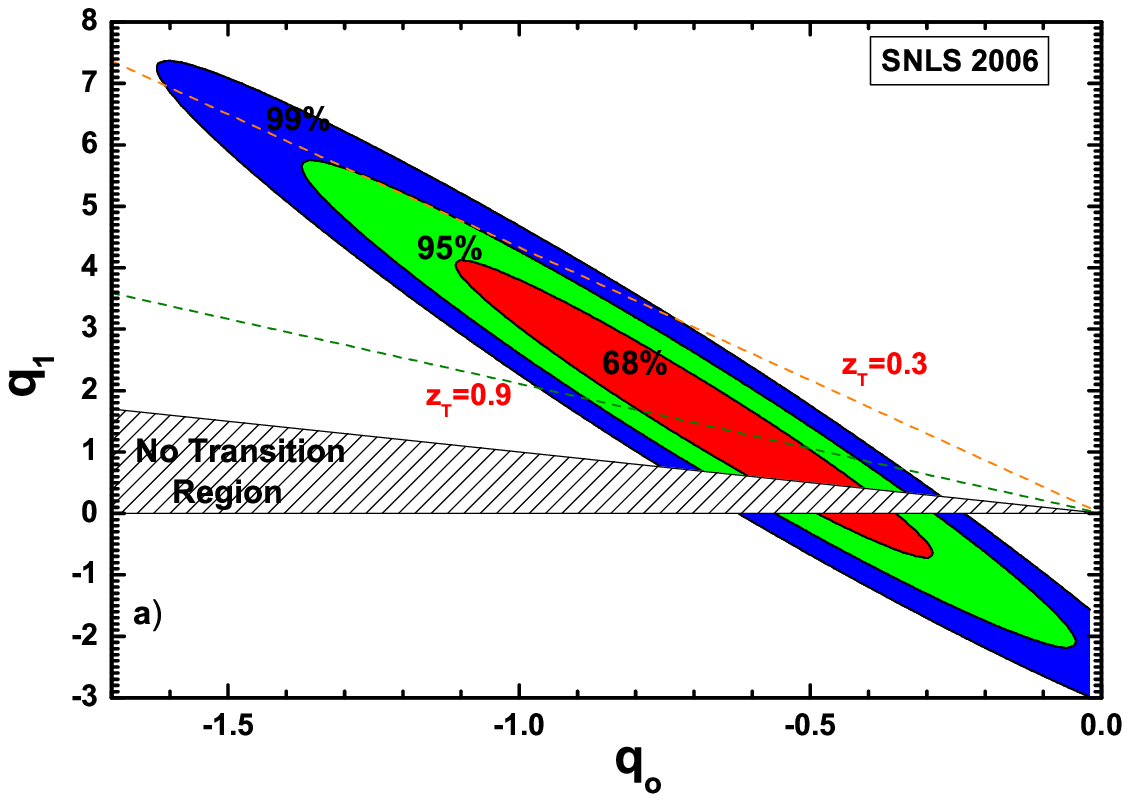}
\includegraphics[width=80mm, angle=0]{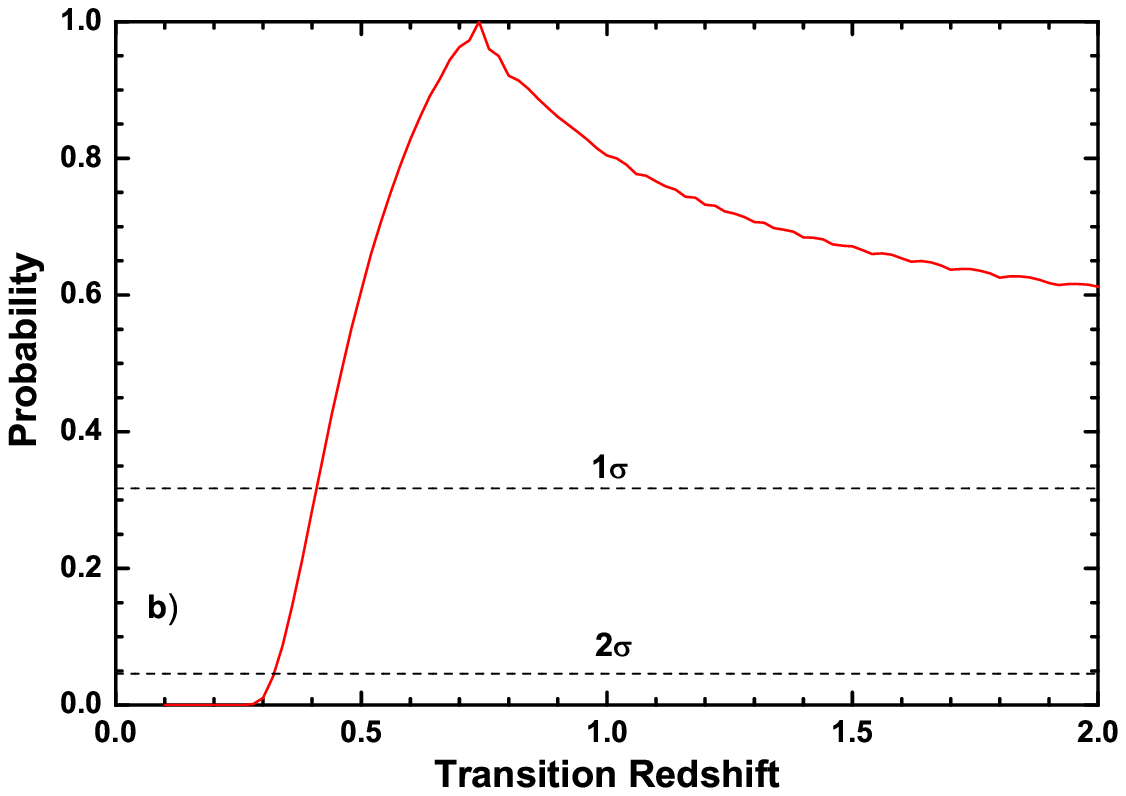}
\caption{{\bf a)} Constraints on the deceleration parameter derived
from the SNLS sample for $q(z)=q_o+q_1z/(1+z)$. Again, the data
favor the recent acceleration and past deceleration. {\bf b)}
Likelihood function for the past transition redshift. Note that the
peak of the likelihood function is still pronounced but the upper
limit of the transition redshift is not very well defined (confront
with Figs. 3b and 4b).} \label{Fig5}
\end{figure}

\subsection{Analysis from SNLS:  $q=q_o + q_1{z}$}

Let us now consider the Astier {\it {et al.}} 2006 data sample to
constrain the plane $q_o-q_1$ for the above linear phenomenological
description.

In figure 4a we show  the plane $q_o-q_1$ with confidence levels of
$68.3\%$, $95.4\%$ and $99.7\%$ to the SNLS data set. With one free
parameter the confidence region is $-1.17 \leq q_o \leq -0.16$ and
$-1.2 \leq q_1 \leq 3.4$ with ($95.4\%$). In the panel we see that
$z_t= 0.3$ is only marginally compatible at 3$\sigma$ while $z_t=
2.0$ is well inside with reasonable confidence (2$\sigma$).
Probably, this apparent conflict comes from the fact that there are
few data with high dispersions in $z\geq 0.8$ (see commentary by
Astier {\it {et al.}} in their paper). Interestingly, the forbidden
region now crosses the confidence level curves (even at 1$\sigma$),
however, the SNLS data (different from the \emph{Gold}07 sample
data) are compatible with the existence of a transition redshifts in
the future, that is, beyond the forbidden region ($z > -1$). In
particular, such a result reinforces the interest to examine higher
order corrections (beyond the first order expansion) within the
kinematic approach.

In figure 4b we show the likelihood function for the transition
redshift. Our analysis furnishes the best fit $z_t =
0.61^{+3.68}_{-0.21}$ ($1\sigma$) for one free parameter. The
likelihood is not well behaved at high redshifts because its upper
limit is not known. Actually, although supplying an excellent
adjustment when combined with the WMAP 3-years and cosmic
concordance model, the superior limit of $z_t$ is not tightly
constrained by the present SNLS sample. This poor inference should
be expected from the very beginning in virtue of the absence of data
for $z>1$ in the SNLS sample.

\subsection{SNLS:  $q=q_o + q_1{z}/{1 + z}$}

In figure 5a we show  the plane $q_o-q_1$  to the SNLS data set in
the context  of the above quoted parameterization. The allowed
regions at $68\%$, $95\%$ and $99\%$ are indicated. With one free
parameter the confidence region is $-1.24 \leq q_o \leq -0.16$ and
$-1.5 \leq q_1 \leq 4.9$ with ($95.4\%$). In the panel we see that
$z_t= 0.3$ (flat $\Lambda$CDM with $\Omega_m\simeq 0.5$) is outside
of the allowed (2$\sigma$) region  while $z_t= 2.0$ (flat
$\Lambda$CDM with $\Omega_m\simeq 0.07$ and $\Omega_{\Lambda}\simeq
0.93$) is well inside.

It is worth noticing that the forbidden region is now represented by
a right-angle triangle. Interestingly, it crosses the confidence
level curves (even at 1$\sigma$), however, the SNLS data (different
from the \emph{Gold}07 sample data) are compatible with the
existence of a transition redshifts in the future, that is, for
negative value of $z$. In particular, such a result reinforces the
interest to examine higher order corrections (beyond the first order
expansion) within the kinematic approach. In particular,  this means
that some theoretical problems associated with the existence of
horizons in the future of  models that accelerates forever must be
solved in a natural way (see, for instance, Carvalho {\it et al.}
(2006)).

In figure 5b we show the likelihood function of the transition
redshift. Our analysis furnishes the best fit $z_t = 0.74$ with
inferior value $0.42$ ($1\sigma$) for one free parameter and the
upper value to $z_t$ is too high. Actually, it is very high and
completely incompatible with observations (see also Fig. 4b). Such a
poor inference at high $z$ from  SNLS sample (Astier {\it{et al.}}
2006) for  both parameterizations should be expected  because the
absence of data for $z>1$. This is calling our attention to the
importance of observing high redshift supernovae within the SNLS
collaboration.

\begin{figure}[ht]
\centering
\includegraphics[width=80mm, angle=0]{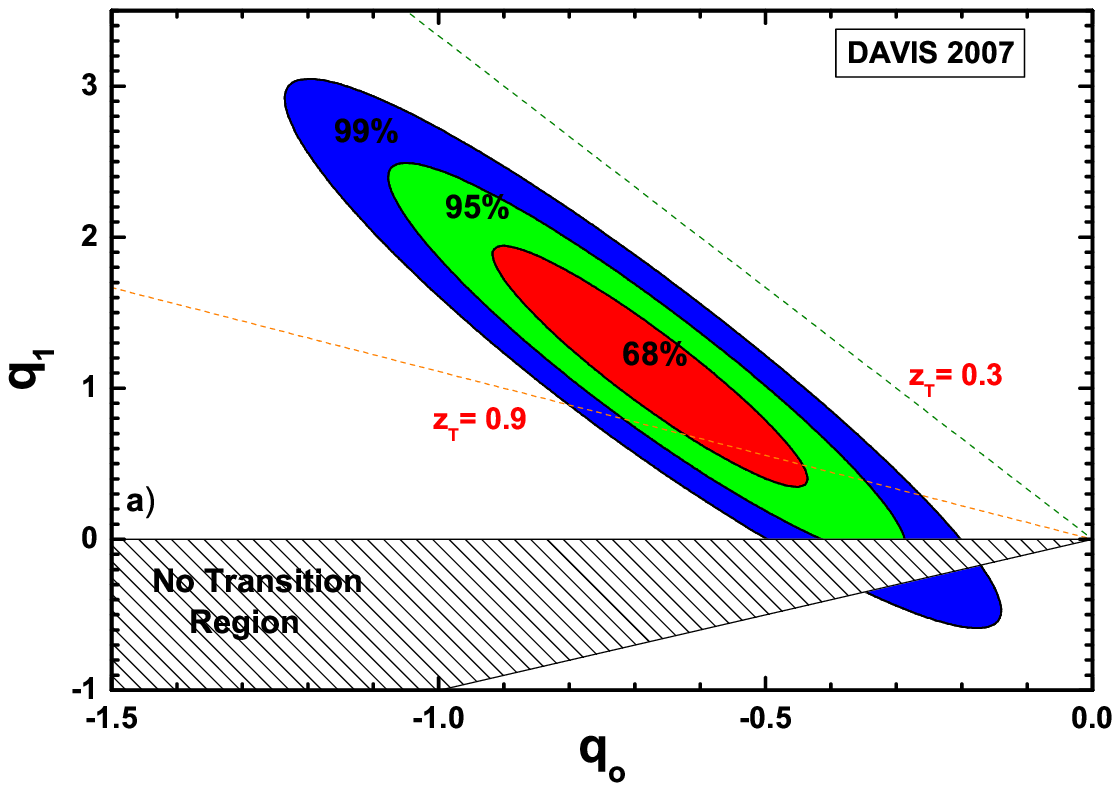}
\includegraphics[width=80mm, angle=0]{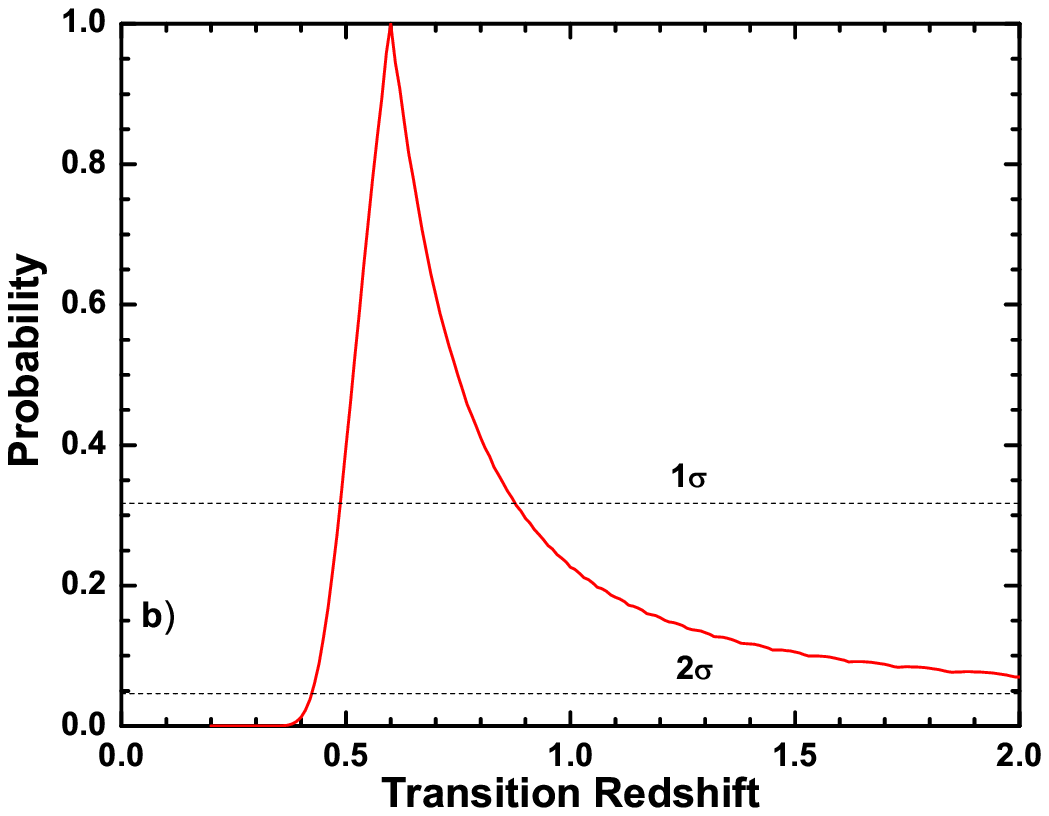}
\caption{{\bf a)} Constraints for the deceleration parameter are
derived from the Davis {\it{et al.}} 2007 analysis for
$q(z)=q_o+q_1z$. 68.3\%, 95.4\% and 99.7\% confidence intervals are
showed in the graph. {\bf b)} Likelihood function for the past
transition redshift for a two-parameter model of the expansion
history. Our analysis furnishes the best fit $z_t =
0.60^{+0.28}_{-0.11}$ ($1\sigma$).} \label{Fig6}
\end{figure}

\begin{figure}[ht]
\centering
\includegraphics[width=80mm, angle=0]{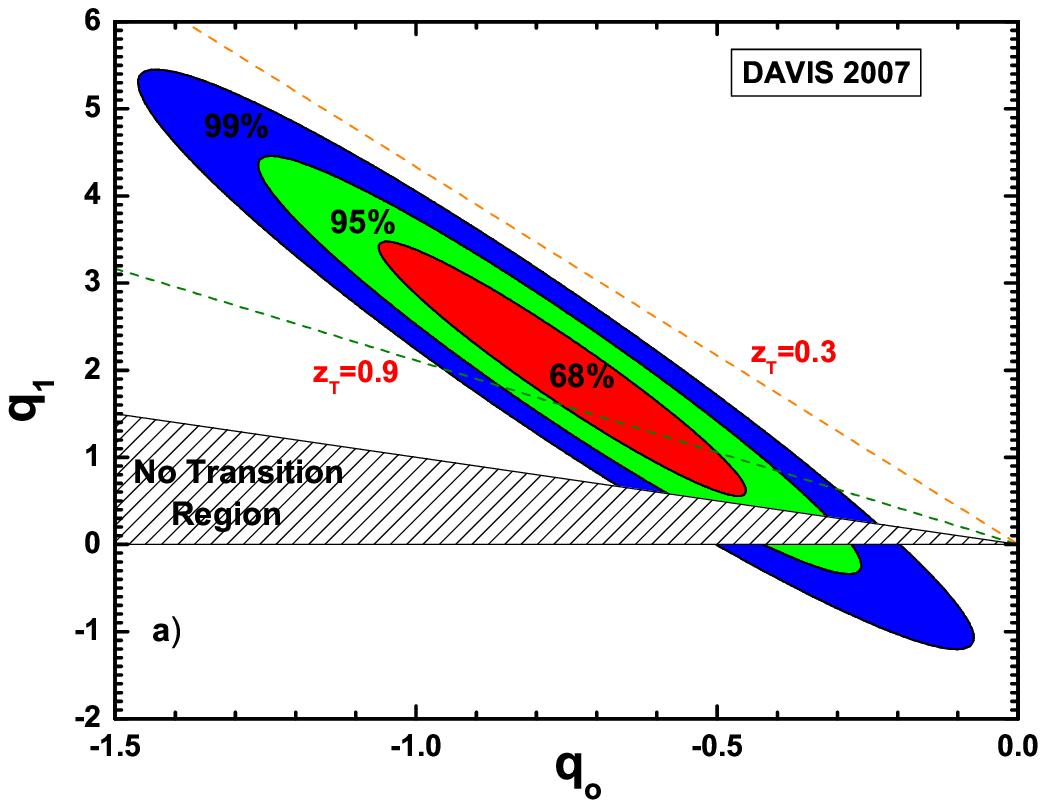}
\includegraphics[width=80mm, angle=0]{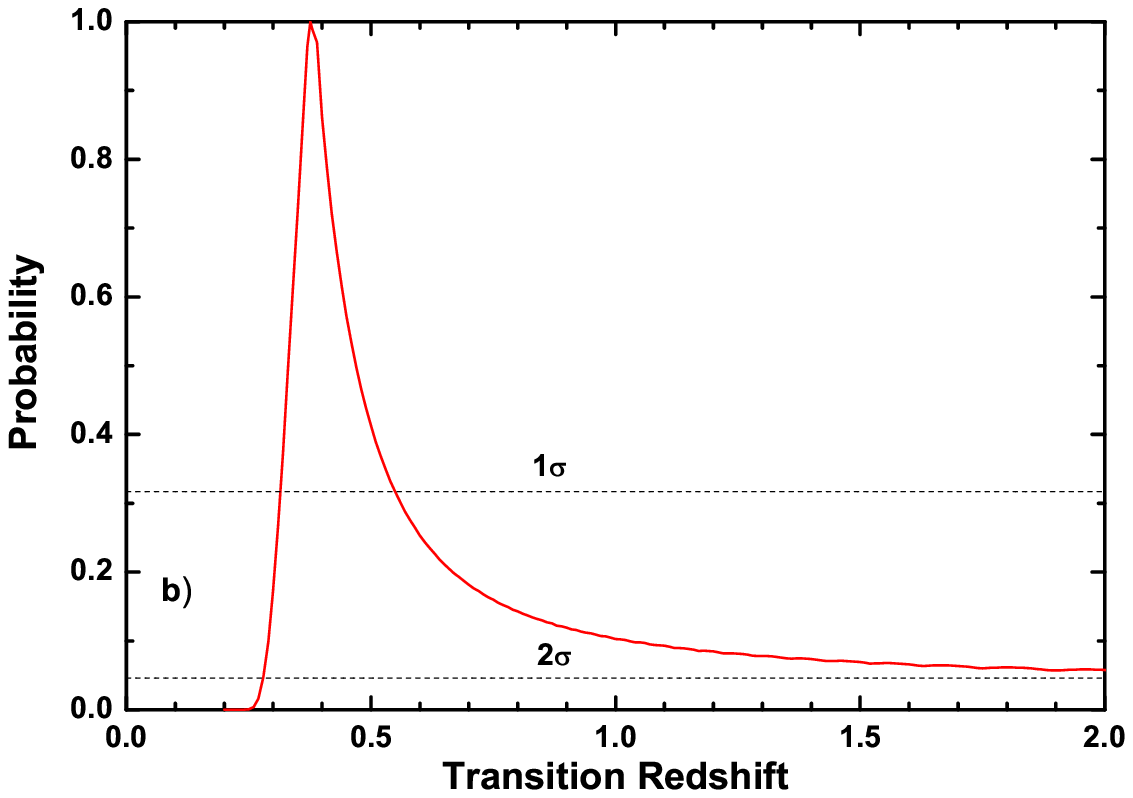}
\caption{{\bf a)} Constraints on the deceleration parameter are
derived from the Davis {\it{et al.}} 2007 analysis. The data favor
the recent acceleration ($q_0<0$) and past deceleration ($q_1>0$)
with high confidence for $q(z)=q_o+q_1z/(1+z)$. {\bf b)} Likelihood
function for the past transition redshift for a two-parameter model
of the expansion history, $q(z)=q_o+q_1z/(1+z)$, from SNe Ia. Our
analysis furnishes an accurate best fit $z_t = 0.38^{+0.17}_{-0.06}$
($1\sigma$).} \label{Fig7}
\end{figure}

\subsection{Davis 2007 Sample:  $q=q_o + q_1{z}$}

Let us now discuss the constraints within the linear approach for
the Davis {\it{et al.}} data set. In figure 6a we display the plane
$q_o-q_1$ for confidence levels of $68.3\%$, $95.4\%$ and $99.7\%$
as indicated. Note that such a sample also favors a Universe with
recent acceleration ($q_o<0$) and previous deceleration ($dq/dz>0$).
With one free parameter the confidence region is $-1.0 \leq q_o \leq
-0.36$ and $0.11 \leq q_1 \leq 2.2$ with ($95.4\%$). In the same
panel we also see that $z_t= 0.3$ is completely outside even of the
significant $3\sigma$ region while $z_t= 0.9$ is well inside with
high confidence.

In figure 6b we show the likelihood function for the transition
redshift. We can see that $z_t=0.60^{+0.28}_{-0.11}$ ($1\sigma$) for
one free parameter. These results are in better accord with flat
LCDM models. Such a fact may be somewhat related to the higher
degree of statistical completeness obtained when the SNe type Ia
samples are added.

\subsection{Davis 2007 Sample:  $q=q_o + q_1{z}/{1 + z}$}

In Figure 7a we display the plane $ q_o - q_1$ when Davis sample is
considered for $q(z)=q_o+q_1z/(1+z)$ parameterization. With one free
parameter the confidence region is $-1.17 \leq q_o \leq -0.36$ and
$0.39 \leq q_1 \leq 3.99$ with ($95.4\%$). In the same panel we also
see that $z_t= 0.3$ is outside while $z_t= 0.9$ is well inside with
high confidence.

\begin{figure}[ht]
\centering
\includegraphics[width=80mm, angle=0]{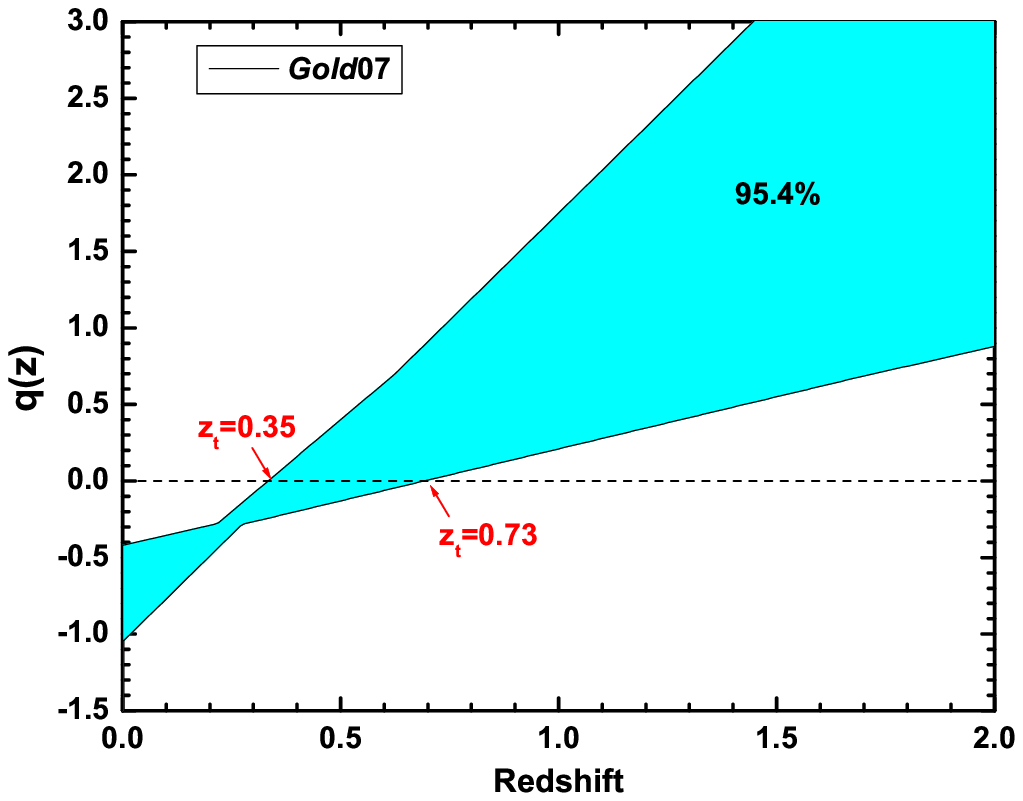}
\includegraphics[width=80mm, angle=0]{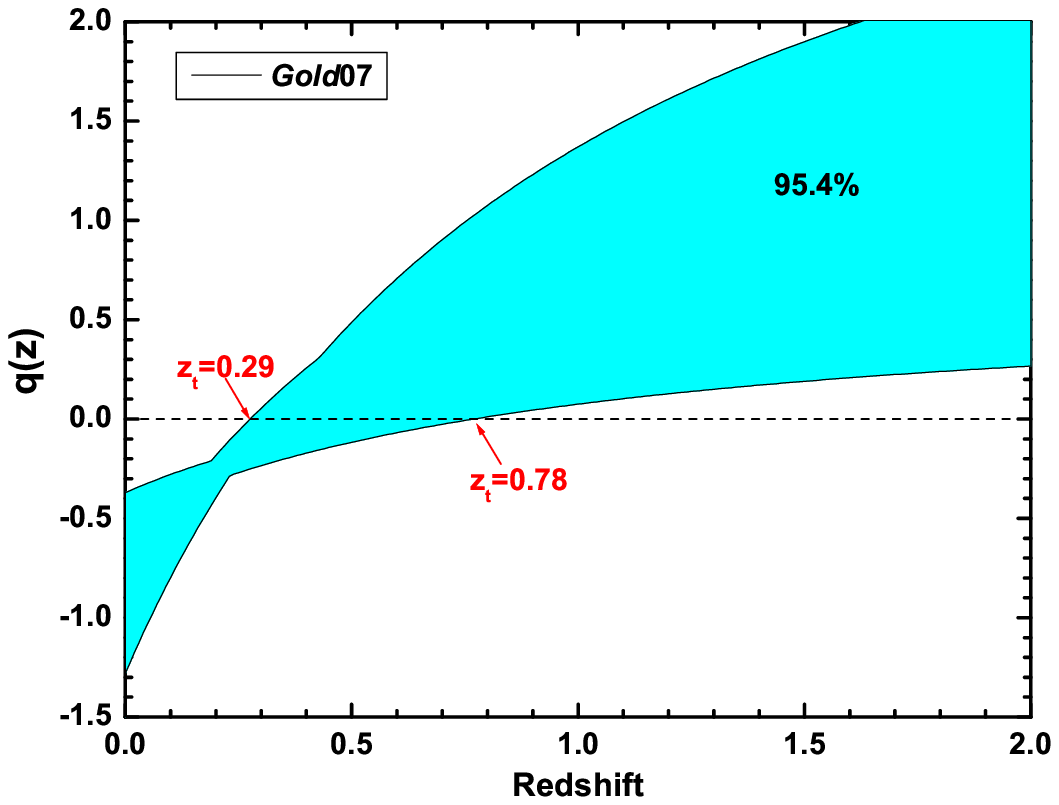}
\caption{{\bf a)} Evolution of the decelerating parameter as a
function of the redshift based on the expansion  $q(z)=q_o+q_1 z$.
{\bf b)} The same of Fig. (a) for a two-parameter expansion history
driven by  $q(z)=q_o+q_1z/(1+z)$.  In the two panels the shadowed
regions means 2$\sigma$ level for the \emph{Gold}07 sample. The
dotted horizontal lines represent the coasting model ($q(z)\equiv
0$). In both parameterizations the data favor a low redshift
accelerating stage of the Universe ($q_0<0$) and a past decelerating
phase ($q_{1} > 0$)  with high confidence level.} \label{Fig8}
\end{figure}

In figure 7b we show the likelihood function of the transition
redshift. We can see that $z_t=0.38^{+0.17}_{-0.06}$ ($1\sigma$) for
one free parameter. This results is comparable with the constraints
from the \emph{Gold07} sample for the same parameterization (see
subsection 2.2). It is also interesting that  the Davis {\it{et
al.}} sample also suggests (for both parameterizations) the
existence of a transition redshift in the future ($z < 0$). Such a
possibility it will be discussed with more  detail in a forthcoming
communication (Lima and Cunha 2008).

\section{RECONSTRUCTING THE DECELERATING PARAMETER}

At this point it is interesting to investigate how a joint analysis
constrains the redshift evolution of the decelerating parameter.
However, since the time evolutions are quite similar for the
different parameterizations, in what follows we focus our attention
on the \emph{Gold}07 sample. The basic results are displayed in
Figures 8a and 8b.

In  Figs. 8a  we see the evolution of the decelerating parameter as
a function of the redshift for the parameterization $q(z)=q_o+q_1z$.
The shadow denotes the 2$\sigma$ region.

In Fig 8b the evolution is shown for a two-parameter expansion
history driven by  $q(z)=q_o+q_1z/(1+z)$. For both cases, we also
show the lower and upper values ($2\sigma$) of the transition
redshifts which is defined by the condition ($q(z)\equiv0$).  Note
also that  for both parameterizations the data favor the recent
acceleration ($q_0<0$) and past deceleration ($q_1>0$) with high
confidence level.

In  Table 1 we summarize the basic constraints on the transition
redshift ($z_t$) established in the present work. As shown there,
for a given phenomenological law the limits were derived separately
for each sample of SNe type Ia data.

\section{CONCLUSION}

The most remarkable dynamical feature distinguishing  dark energy
and matter dominated models  is the present accelerating stage.
However, the observed structures must be formed  in a past
decelerated phase dominated by matter. This leads naturally to a
dynamic phase transition from decelerating to accelerating in the
course of the Universe's evolution. This is the most natural
scenario within the framework of the general relativity. It has been
confirmed by several disparate observations ranging from the
anisotropy of the CMB, SNe Ia, galaxy clusters, age of the universe,
however, even considering the high-quality of the present data the
whole set is still compatible with many different dark energy
models.

The basic information about the deceleration parameter has been
obtained here through a model independent analysis using two
different phenomenological laws, namely: $q(z) = q_o + q_1z$, as
proposed by Riess {\it et al.} (2004) and $q(z)=q_o+q_1z/(1+z)$ (Xu
{\it{et al.}} 2007). The contours to the plane ($q_o, q_1$) and the
likelihood function for transition redshift were discussed in detail
with basis on the data sample by Riess {\it{et al.}} (2007),  Astier
{\it{et al.}} (2006) and Davis {\it{et al.}} (2007). In order to
compare the predictions,  we have presented the analysis  first for
each sample separately. As we have seen, although considering the
presence of no transition regions (see Figs. 2a, 3a, 4a, 5a, 6a and
7a), both sample strongly favor a Universe with recent acceleration
($q_o<0$) and previous deceleration ($dq/dz>0$).

The likelihood function for the transition redshift is more well
defined for the Riess {\it{et al.}} (2007) and Davis {\it{et al.}}
(2007) data. In table 1 we summarize the main results related to the
transition redshift $z_t$ in the present kinematic approach. For
both samples, our analysis provides a model independent evidence
that a dynamic phase transition from decelerating to accelerating
happened at redshifts smaller than unity. Hopefully, the constraints
on $z_t$ will be considerably improved in the near future with the
increasing of supernovae data at intermediate and high redshifts.

\begin{deluxetable}{lccc}
\tablewidth{0pt} \tabletypesize{\footnotesize} \tablecaption{Limits
on the transition redshift $z_t$\label{Tab2}}
\tablehead{{\hspace{1.5cm}} \\
\colhead{Sample} & \colhead{Best fit} & \colhead{Confidence region
($1\sigma$)}} \startdata
 & $q=q_o + q_1{z}$ & \\ \hline
Riess07..........& $0.43$ & $0.38\leq z_t \leq 0.52$\\
Astier06.........& $0.61$ & $0.40\leq z_t \leq 4.29$ \\
Davis07.........& $0.60$ & $0.49\leq z_t \leq 0.88$ \\
\hline
 & $q=q_o + q_1{z}/{(1 + z)}$ & \\ \hline
Riess07..........& $0.39$ & $0.33\leq z_t \leq 0.49$\\
Astier06.........& $0.74$ & $0.42\leq z_t $ high \\
Davis07.........& $0.38$ & $0.32\leq z_t \leq 0.55$ \\ \hline
\enddata
\end{deluxetable}

\section*{Acknowledgments}
The authors are grateful to R. C. Santos, L. Marassi and S. Vitenti
for helpful discussions. This work was partially supported by the
CNPq - Brazil and FAPESP. JASL is grateful to FAPESP grant No.
04/13668 and JVC is supported by FAPESP No. 05/02809-5.

{}

\section*{Appendix}

\appendix

\section{Distance Luminosity in the Kinematic Description}
\label{dataan}

In this Appendix, we shall outline the method used to obtain the
theoretical expression for $D_{L}(z)$ when the kinematic approach is
based on the expansion: $q(z)= q_o + q_1z$. The result is valid for
a  geometrically  flat model. By definition:

\begin{equation}\label{eq:dLq1}
D_L (z) = (1+z)\int_0^z\, \frac{du}{H(u)}
\end{equation}
where we have set $c=1$ and H is the Hubble parameter which in the
kinematic approach must be obtained in terms of the decelerating
parameter:
\begin{equation}\label{qz1}
q(z)\equiv -\frac{a\ddot a}{\dot a^2} = \frac{d H^{-1}(z)}{ dt} -1 =
{-\frac{a}{H}}\frac{dH}{da}-1
\end{equation}
By integrating the above equation and inserting the result into
(\ref{eq:dLq1}) it follows that
\begin{equation}\label{eq:dLq3}
D_L(z)=\frac{(1+z)}{H_0}\int_0^z\, \exp{\left[-\int_0^u\,
[1+q(u)]d\ln (1+u)\right]}\, du
\end{equation}
which is just (\ref{eq:dLq}). Replacing the expansion of $q(z)$, we
obtain
\begin{equation}
\label{eq:Ho2} D_{L}(z) = (1+z)H_o^{-1} \int_0^z (1 +
u)^{\nu-1}\exp\left[{-\mu u}\right]\,du
\end{equation}
where $\mu=q_1$ and $\nu=q_1-q_o$. Now, a simple variable change
results
\begin{eqnarray}\label{eq:dLKin3}
D_L (z) &=& \frac{(1+z)}{H_{0}}e^{q_{1}}q_{1}^{q_{o}-q_{1}}
[\gamma(q_{1}-q_{o},(1+z)q_1) \nonumber \\
& & \,\, - \gamma{({q_1-q_o},{q_1})}],
\end{eqnarray}
where $\gamma$ is the incomplete gamma function (Abramowitz \&
Stegun 1972).

Interestingly, in the linear approximation the luminosity distance
can also be directly written in terms of the transition redshift:
\begin{eqnarray}\label{eq:dLKin4}
D_L (z) = \frac{(1+z)}{H_0} e^{|q_o|\over z_t}
{{\left(\frac{|q_o|}{z_t}\right)}}^{q_o(1+1/z_t)}\nonumber\\
\,\left(\gamma\left[{|q_o|(1 + 1/z_t)},{|q_o|\over z_t}(1 +
z)\right]-\gamma{\left[|q_o|(1 + 1/z_t), {|q_o|\over
z_t}\right]}\right).
\end{eqnarray}

For the second parametrization $q(z)=q_o+q_1z/(1+z)$, we obtain
\begin{equation}
\label{eq:Ho2Paramet} D_{L}(z) = (1+z)H_o^{-1} \int_0^z (1 +
u)^{\nu-1}\exp\left[{-u\mu}\right]\,du
\end{equation}
where $\mu=q_1$ and $\nu=q_1+q_o$. As well as in the last integral,
a simple variable change results
\begin{eqnarray}\label{eq:dLKin3Paramet}
D_L (z) &=& \frac{(1+z)}{H_{0}}e^{q_{1}}{q_1^{-(q_o+q_1)}}
[\gamma(q_{1}+q_{o},q_1) \nonumber \\
& & \,\, - \gamma{({q_1+q_o},q_1/(1+z))}].
\end{eqnarray}

Or yet, in terms of the transition redshift:
\begin{eqnarray}\label{eq:dLKin4Paramet}
D_L (z) = \frac{(1+z)}{H_0} e^{|q_o|(1+z_t)\over z_t}
{{\left(\frac{|q_o|(1+z_t)}{z_t}\right)}}^{q_o/z_t}\nonumber\\
\,\left(\gamma\left[{|q_o|/z_t},{|q_o|(1+z_t)\over
z_t}\right]-\gamma{\left[|q_o|/z_t, {|q_o|(1+z_t)\over z_t(1 +
z)}\right]}\right).
\end{eqnarray}


\begin{thebibliography}{99}
\bibitem[\protect\citeauthoryear{Abramow}{1972}]{b1} Abramowitz M. \& Stegun I. A., 1972,
Handbook of Mathematical Functions. Dover Publications
\bibitem[\protect\citeauthoryear{alcaniz1}{1999}]{a1} Alcaniz J. S. \& Lima  J. A. S. 1999, ApJ, 521, L87, [astro-ph/9902298]
\bibitem[\protect\citeauthoryear{alcaniz1}{2001}]{a1} Alcaniz J. S. \& Lima J. A. S., 2001, ApJ, 550, L133, [astro-ph/0101544]
\bibitem[\protect\citeauthoryear{alcaniz2}{2003}]{a2} Alcaniz J. S., Lima J. A. S. \& Cunha
J. V., 2003, Mon. Not. Roy. Astron. Soc., 340, L39, [astro-ph/0301226]
\bibitem[\protect\citeauthoryear{alcaniz4}{2005}]{a4} Alcaniz J. S. \& Lima J. A. S., 2005,
Phys. Rev. D, 72, 063516, [astro-ph/0507372]
\bibitem[\protect\citeauthoryear{Bent1}{2006}]{b5} Astier P. {\it{et
al.}}, 2006, Astron. Astrophys., 447, 31
\bibitem[\protect\citeauthoryear{Bent1}{2002}]{b5} Bento M. C., Bertolami O. \& Sen A. A.,
2002, Phys. Rev. D, 66, 043507
\bibitem[\protect\citeauthoryear{Bent2}{2003}]{b6} Bento M. C., Bertolami O. \& Sen A.
A., 2003, Phys. Rev. D, 67, 063003
\bibitem[\protect\citeauthoryear{Cald1}{1998}]{b16} Caldwell R. R., Dave R. \& Steinhardt P. J.,
1998, Phys. Rev. Lett., 80, 1582
\bibitem[\protect\citeauthoryear{Cald2}{2000}]{b17} Caldwell R. R., 2000, Braz. J. Phys., 30, 215
\bibitem[\protect\citeauthoryear{Cald3}{2002}]{b16} Caldwell R. R., 2002, Phys. Lett. B, 545, 23
\bibitem[\protect\citeauthoryear{CL205}{2005}]{b1} Carneiro, S. \&  Lima J. A. S., 2005, Int. J. Mod. Phys. A, 20, 2465, [gr-qc/0405141]
\bibitem[\protect\citeauthoryear{carv}{1992}]{b1} Carvalho J. C., Lima J. A. S. \& Waga I.,
1992, Phys. Rev. D, 46, 2404
\bibitem[\protect\citeauthoryear{carv}{1992}]{b1} Carvalho F. C. {\it {et al.}}, 2006, Phys. Rev. Lett. {97}, 081301, [astro-ph/0608439]
\bibitem[\protect\citeauthoryear{chiba}{1997}]{b3} Chiba T., Sugiyama N., and  Nakamura T.,
1997, MNRAS, 289, L5
\bibitem[\protect\citeauthoryear{Chim}{2001}]{b15} Chimento L. P., Jakubi A. S. \& Zuccala
 N. A., 2001, Phys. Rev. D, 63, 103508
\bibitem[\protect\citeauthoryear{Copeland}{2006}]{b16} Copeland E. J., Sami M. \& Tsujikawa S.,
2006, IJMPD, 15, 1753
\bibitem[\protect\citeauthoryear{cunha02}{2002}]{c1} Cunha J. V., Lima J. A. S. \& Alcaniz J.
S., 2002a, Phys. Rev. D, 66, 023520, [astro-ph/0202260]
\bibitem[\protect\citeauthoryear{cunha02b}{2002}]{c1} Cunha J. V., Lima J. A. S. \&
Pires N., 2002b, Astron. \& Astrophys., 390, 809,
[astro-ph/0202217]
\bibitem[\protect\citeauthoryear{cunha04a}{2004}]{c3} Cunha J. V. \& Santos R. C., 2004, Int. J. Mod. Phys. D, 13, 1321, [astro-ph/0402169]
\bibitem[\protect\citeauthoryear{cunha04b}{2004}]{c3} Cunha J. V., Alcaniz J. S. \& Lima J. A.
S., 2004, Phys. Rev. D, 69, 083501, [astro-ph/0306319]
\bibitem[\protect\citeauthoryear{cunha07}{2007}]{c6} Cunha J. V., Marassi L. \& Santos R. C.,
2007, Int. J. Mod. Phys. D, 6, 403, [astro-ph/0608686]
\bibitem[\protect\citeauthoryear{Davis07}{2007}]{Davis07} Davis T. M., M\"{o}rtsell E.,
Sollerman J. {\it{et al.}}, 2007, ApJ, 666, 716
\bibitem[\protect\citeauthoryear{Dev}{1995}]{b16} Dev A., Jain D. \& Alcaniz J. S., 2003,
Phys. Rev. D, 67, 023515
\bibitem[\protect\citeauthoryear{Efst}{1999}]{b16} Efstathiou G., 1999, Mon. Not. Roy.
Astron. Soc., 310, 842
\bibitem[\protect\citeauthoryear{Fabris07}{2007}]{f07} Fabris J. C., Shapiro I. L. \& Sola J.,
2007, J. Cosmol. Astropart. Phys., 0702, 016
\bibitem[\protect\citeauthoryear{Free1}{1987}]{b16} Freese K. {\it{et al.}}, 1987, Nucl. Phys.
B, 287, 797
\bibitem[\protect\citeauthoryear{Free2}{2002}]{b15} Freese K. \& Lewis M., 2002, Phys. Lett. B,
540, 1
\bibitem[\protect\citeauthoryear{Fr95}{1995}]{b16} Frieman J. A. {\it{et al.}}, 1995, Phys. Rev.
Lett., 75, 2077
\bibitem[\protect\citeauthoryear{Gard}{2005}]{b16} Gardner C. L., 2005, Nucl. Phys. B, 707,
278
\bibitem[\protect\citeauthoryear{Gong}{2006}]{b16} Gong Y. \& Wang A., 2006, Phys. Rev. D, 73,
083506
\bibitem[\protect\citeauthoryear{Gong}{2007}]{b17} Gong Y. \& Wang A., 2007, Phys. Rev. D, 75,
043520
\bibitem[\protect\citeauthoryear{Gon04}{2004}]{b16} Gonzalez-Diaz P. F. \& Siguenza C. L.,
2004, Nucl. Phys. B, 697, 363
\bibitem[\protect\citeauthoryear{Ishida08}{2008}]{b16} Ishida E. E. O.,  Reis R. R. R., Toribio  A. V. \& Waga I.,
2008, Astropart. Phys., 28, 547
\bibitem[\protect\citeauthoryear{Komatsu2008}{2001}]{b16} Komatsu E. {\it et al.}, 2008 ({ArXiv:
0803.0547v1})
\bibitem[\protect\citeauthoryear{Ozer}{2001}]{b16} Kamenshchik A., Moschella U. \& Pasquier V.,
2001, Phys. Lett. B, 511, 265
\bibitem[\protect\citeauthoryear{perm1}{2008}]{b16} Kowalski M., {\it et al.}, 2008
({ArXiv:08044142v1})
\bibitem[\protect\citeauthoryear{kujat}{2002}]{b18} Kujat J., Linn A. M., Scherrer R. J.,
Weinberg D. H., 2002, ApJ, 572, 1
\bibitem[\protect\citeauthoryear{Jassal}{2006}]{b18} Jassal H. K., Bagla J. S. \& Padmanabhan
T., 2006 (astro-ph/0601389)
\bibitem[\protect\citeauthoryear{LimaMaia99}{2006}]{ML06} Lima J. A. S. \& Maia J. M. F., 1993, Mod. Phys. Lett. A, 8, 591

\bibitem[\protect\citeauthoryear{LimaMaia94}{2006}]{ML06} Lima J. A. S. \& Maia J. M. F., 1994, Phys. Rev. D, 49, 5597

\bibitem[\protect\citeauthoryear{Lima}{1996}]{b16} Lima J. A. S., 1996, Phys. Rev. D, 54, 2571, [gr-qc/9605055]

\bibitem[\protect\citeauthoryear{Lima1}{1999}]{b15} Lima J. A. S. \& Alcaniz J. S., 1999, Astron.
Astrophys, 348, 1, [astro-ph/9902337]
\bibitem[\protect\citeauthoryear{LimCunAl03}{2003}]{Li03} Lima J. A. S., Cunha J. V. \& Alcaniz
J. S. 2003, Phys. Rev. D, 68, 023510, [astro-ph/0303388]
\bibitem[\protect\citeauthoryear{Lima}{2004}]{b16} Lima J. A. S., 2004, Braz. J.
Phys., 34, 194, [astro-ph/0402109]
\bibitem[\protect\citeauthoryear{LimaAlc04}{2004}]{LA04} Lima J. A. S. \& Alcaniz J. S., 2004,
Phys. Lett. B, 600, 191,  [astro-ph/0402265]
\bibitem[\protect\citeauthoryear{LimCunAl06}{2006}]{Li06} Lima J. A. S., Cunha J. V. \& Alcaniz
J. S. 2006a, [astro-ph/0608469]
\bibitem[\protect\citeauthoryear{LimCunAl06b}{2006}]{Li06} Lima J. A. S., Cunha J. V. \& Alcaniz J. S. 2006b,
[astro-ph/0611007]

\bibitem[\protect\citeauthoryear{MaiaLima99}{2006}]{ML06} Maia J. M. F. \& Lima J. A. S., 1999, Phys. Rev. D, 60, 101301, [astro-ph/9910568]

\bibitem[\protect\citeauthoryear{MaiaLima02}{2006}]{ML06} Maia J. M. F. \& Lima J. A. S., 2002, Phys. Rev. D, 65, 083513, [astro-ph/0112091]

\bibitem[\protect\citeauthoryear{OvCo}{1998}]{b16} Overduin F. M. \& Cooperstock F. I.,
1998, Phys. Rev. D, 58, 043506
\bibitem[\protect\citeauthoryear{Ozer}{1987}]{b16} Ozer M. \& Taha O., 1987, Nucl. Phys. B,
287, 797
\bibitem[\protect\citeauthoryear{Padman}{2003}]{b26} Padmanabhan T., 2003, Phys. Rep., 380, 235
\bibitem[\protect\citeauthoryear{Pavon}{2003}]{b26} Pavon D. \& Zimdahl W., 2005, Phys. Lett. B628, 206
\bibitem[\protect\citeauthoryear{Peebles}{2003}]{b16} Peebles P. J. E., 1993, Principles of Physical Cosmology, Princeton UP, New Jersey
\bibitem[\protect\citeauthoryear{Peebles}{2003}]{b16} Peebles P. J. E. \& Ratra B., 2003, Rev.
Mod. Phys., 75, 559
\bibitem[\protect\citeauthoryear{Perivo05}{2005}]{Per1} Perivolaropoulos
L., 2005, Phys. Rev. D, 71, 063503
\bibitem[\protect\citeauthoryear{perm1}{1999}]{b16} Perlmutter S., {\it et al.}, 1999, ApJ,
517, 565
\bibitem[\protect\citeauthoryear{Qiang}{2006}]{b16} Qiang Y. \& Zhang T.-J., 2006, Mod. Phys.
Lett. A, 21, 75
\bibitem[\protect\citeauthoryear{ratra}{1988}]{b212} Ratra B. \& Peebles P. J. E., 1988,
Phys. Rev. D, 37, 3406
\bibitem[\protect\citeauthoryear{riess}{1998}]{b21} Riess A. G. {\it et al.}, 1998, AJ, 116,
1009
\bibitem[\protect\citeauthoryear{riess}{2004}]{b21} Riess A. G. {\it et al.}, 2004, ApJ, 607,
665
\bibitem[\protect\citeauthoryear{riess}{2007}]{b21} Riess A. G. {\it et
al.}, 2007, ApJ, 659, 98
\bibitem[\protect\citeauthoryear{Sain}{2000}]{b16} Saini T. D. {\it{et al.}}, 2000, Phys.
Rev. Lett., 85, 1162
\bibitem[\protect\citeauthoryear{Santos}{2008}]{b1} Santos R. C., Cunha J. V. \& Lima J. A. S., 2008, Phys. Rev. D 77, 023519, arXiv:0709.3679 [astro-ph]
\bibitem[\protect\citeauthoryear{Santos}{2008}]{b1} Santos R. C. \& Lima, J. A. S., 2008, Phys. Rev. D 77, 083505, arXiv:0803.1865 [astro-ph]
\bibitem[\protect\citeauthoryear{Sain}{2000}]{b16} Spergel D. N. {\it{et al.}}, 2007, ApJ, 170, 377
\bibitem[\protect\citeauthoryear{turner}{1997}]{b15} Turner M. S. \& White M., 1997, Phys.
Rev. D, 56, R4439
\bibitem[\protect\citeauthoryear{turner}{2002}]{b15} Turner M. S. \& Riess A. G., 2002, ApJ,
569, 18
\bibitem[\protect\citeauthoryear{Virey}{2005}]{b15} Virey J.-M. {\it{et al.}}, 2005, Phys.
Rev. D, 72, R061302
\bibitem[\protect\citeauthoryear{Weinberg}{1972}]{We72} Weinberg S., 1972, Cosmology and
Gravitation, John Wiley Sons, N.Y.
\bibitem[\protect\citeauthoryear{Wood-Vasey07}{2007}]{Wo07} Wood-Vasey W. M., Miknaitis G., Stubbs C.
W., 2007, ApJ, 666, 694
\bibitem[\protect\citeauthoryear{Xu}{2007}]{Xu07} Xu L., Zhang C., Chang B. \& Liu H., 2007
(astro-ph/0701519)


\end{thebibliography}
\end{document}